\documentclass[prl,reprint, nofootinbib,amsmath,amssymb, aps, floatfix]{revtex4-1}
\usepackage{graphicx}%, xcolor, varwidth}
\usepackage{color}
\usepackage{braket}
\usepackage{mathrsfs}
\usepackage{amsthm}
\newtheorem{thm}{Theorem}
\newtheorem{cor}[thm]{Corollary} 
\newtheorem{lem}[thm]{Lemma}
\newtheorem{prop}[thm]{Proposition}
\theoremstyle{remark}
\newtheorem{defn}[thm]{Definition}

\newtheorem{conj}[thm]{Conjecture}
\newtheorem{conv}[thm]{Convention}

\DeclareMathOperator{\A}{Area}

\usepackage{bbold}
\usepackage{comment}

\begin{document}
\title{Singularities From Entropy}
\author{Raphael Bousso}
\affiliation{Berkeley Center for Theoretical Physics and Department of Physics,\\
University of California, Berkeley, CA 94720, USA} %and \\
%Lawrence Berkeley National Laboratory, Berkeley, CA 94720, U.S.A.} 
\author{Arvin Shahbazi-Moghaddam}
\affiliation{Stanford Institute for Theoretical Physics,\\ Stanford University, Stanford, CA 94305, USA}

\begin{abstract}
Assuming the Bousso bound, we prove a singularity theorem: if the light rays entering a hyperentropic region contract, then at least one light ray must be incomplete. ``Hyperentropic’' means that the entropy of the region exceeds the Bekenstein-Hawking entropy of its spatial boundary. Our theorem provides a direct link between singularities and quantum information. The hyperentropic condition replaces the noncompactness assumption in Penrose’s theorem, so our theorem is applicable even in a closed universe. In an asymptotically de Sitter spacetime, for example, a big bang singularity can be diagnosed from the presence of dilute radiation at arbitrarily late times. In asymptotically flat space, Penrose’s theorem can be recovered by adding soft radiation.

\end{abstract}
\maketitle

\section{Introduction} \label{sec-intro}

Penrose's singularity theorem~\cite{Penrose:1964wq} is of fundamental importance to general relativity and quantum gravity. It implies that singularities are generic and not just an artifact of highly symmetric solutions.

A singularity---defined as the incompleteness of a null or timelike geodesic---signifies a breakdown of classical spacetime. One expects a quantum gravity theory to provide a consistent mathematical description, though it may not be geometric.

The half-century since Penrose's work has seen breakthrough advances in quantum gravity~\cite{Bekenstein:1972tm,Hawking:1974sw, Maldacena:1997re}. Yet, the nature of singularities remains largely a mystery. This is particularly vexing since the Universe contains black holes and may also have a singularity in the past.

It would be helpful, therefore, to gain a new perspective on singularities. In this work, we prove that a singularity must form when the entropy in a spatial region exceeds the maximum entropy allowed on a lightsheet emanating into the same region. Our result suggests that singularities are the response of the spacetime geometry to the presence of too much quantum information.

The Bousso bound~\cite{Bousso:1999xy} states that the renormalized entropy of the quantum fields on a lightsheet $L$ emanating from a surface of area $A$ is bounded by $A/4G$, where $G$ is Newton's constant. A lightsheet is an everywhere nonexpanding null hypersurface. Renormalizing the entropy means stripping off universal divergences associated with entanglement of vacuum modes across the boundary of $L$. (See, \emph{e.g.}, Refs.~\cite{Bousso:1999xy,Bousso:1999cb,Flanagan:1999jp,Bousso:2002ju,Strominger:2003br,Bousso:2003kb,Bousso:2014sda,Bousso:2014uxa} for evidence and generalizations. We set $k_B=c=\hbar=1$.)

By contrast, the entropy in a spatial region $B$ (a partial Cauchy surface) is generally unconstrained by the area of its boundary $\partial B$. The simplest example is a spatially closed universe with nonzero entropy. Or consider a star in an advanced stage of gravitational collapse: its surface area becomes small, but its entropy cannot decrease. We shall call a region \emph{hyperentropic} if its renormalized entropy satisfies $S(B)>A(\partial B)/4G$.

\begin{figure}[b]
\includegraphics[width=0.18\textwidth]{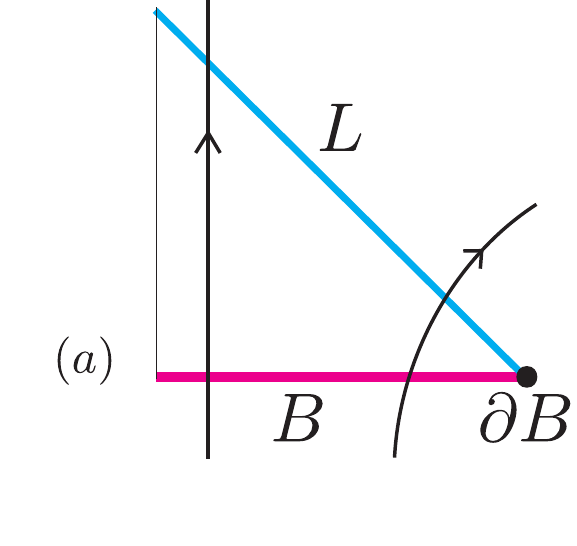}
  \hspace{1cm}
\includegraphics[width=0.18\textwidth]{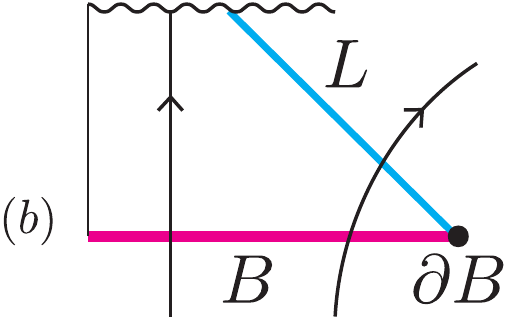}
\caption{\label{fig-closed}
{\footnotesize In these Penrose diagrams, the thin vertical line represents $r=0$; $L$ is a contracting light cone (generally, a lightsheet); and $B$ is a spatial region. (a) If $L$ ``closes off,'' then all matter in $B$ (arrows) must pass through $L$, so its entropy is bounded by the area of $\partial B$. (b) If $B$ is hyperentropic [$S(B)>A(\partial B)/4G$], then $L$ cannot close off, or else the Bousso bound would be violated. We prove that this implies a singularity.}}
\end{figure}

But if $B$ and $L$ have the same domain of dependence,  \emph{i.e.}\ if $D(B)=D(L)$ (see Def.~\ref{ddef} below), then the quantum states on $B$ and $L$ are unitarily equivalent and have the same entropy~\cite{Bousso:1999xy}. In this case, the Bousso bound on $L$ implies that $B$ cannot be hyperentropic; see Fig.~\ref{fig-closed}\emph{a}.

The condition $D(B)=D(L)$ can be broken up as follows: first, the nonexpansion condition must hold on $\partial B$, so that a lightsheet $L$ directed towards the interior of $B$ exists; and secondly, $L$ has to ``close off,'' so that the same matter passes through $B$ and through $L$.

Our central observation is that a contrapositive of this result depends only on initial data on $B$. Suppose that $B$ is hyperentropic, and suppose moreover that a lightsheet $L$ towards the interior of $B$ exists. Then $L$ must fail to close off, or else the Bousso bound would be violated. For example, $L$ might fail to close off if it expands out and runs off to infinity. But since $L$ is an initially contracting null hypersurface, this possibility is excluded~\cite{Bousso:1999cb} by the null energy condition\footnote{This assumption can be eliminated~\cite{QuantumTheorem} by taking a quantum extension of the Bousso bound~\cite{Bousso:2015mna} as the starting point.} (see Def.~\ref{ncc}). Therefore, $L$ must run into a singularity, as shown in Fig.~\ref{fig-closed}\emph{b}.

%In Sec.~\ref{sec-cst} 
We will now make this heuristic argument rigorous and derive a singularity theorem.
%In Sec.~\ref{sec-dis}, 
We will then discuss the relation between Penrose's theorem and ours; and 
%In Sec.~\ref{sec-ex} 
we will illustrate our theorem in several examples where Penrose's theorem does not apply. 

\section{Singularity Theorem} \label{sec-cst}
\begin{conv}\label{Mconv}
Let $(M, g_{ab})$ be a time-orientable globally hyperbolic spacetime. ($M$ may be extendible.) We use $\partial$ to denote a boundary as a subset of a Cauchy slice $\Sigma$ and an overdot to represent a boundary as a subset of $M$.
\end{conv}
\begin{conv}\label{bcconv}
Everywhere below, $B$ will denote a closed subset of a Cauchy slice $\Sigma$ of $M$, such that $\partial B$ is a compact codimension 2 submanifold of $M$ and $B-\partial B\neq \varnothing$.  $C$ denotes the complement of $B$ on $\Sigma$. Thus, $C$ is an open subset of $\Sigma$ and $\partial C = \partial B$.
\end{conv}

\begin{defn}\label{ddef}
For any closed achronal set $K\subset M$, the \emph{future domain of dependence}, $D^+(K)$, is the set of points $p$ such that every past-inextendible causal curve through $p$ must intersect $K$. The \emph{past domain of dependence}, $D^-(K)$, is defined analogously. The \emph{domain of dependence} is $D(K)\equiv D^+(K)\cup D^-(K)$.
\end{defn}

\begin{defn}\label{ijdef}
The chronological and causal future and past, $I^\pm(K)$ and $J^\pm(K)$, of any set $K\subset M$ are defined as in Wald~\cite{wald2010general}. For $K=\set{p}$, it is conventional to drop the set brackets. For convenience we list key consequences of these definitions that will be important below: $p\notin I^+(p)$ but $p\in J^+(p)$; $I^+(K)$ is open; and since we assume global hyperbolicity, for compact $K$, $J^+(K)$ is closed.
\end{defn}

\begin{defn}\label{idef}
For any set $K\subset M$, we define its \emph{domain of influence} as the union of $K$ and all points that can be reached by a timelike curve from $K$:
$I(K) \equiv I^+(K) \cup I^-(K) \cup K$.
\end{defn}

\begin{lem}\label{lem-D+}
With $B$ as in Convention~\ref{bcconv}, $D^\pm(B)$ and $D(B)$ are closed.
\end{lem}

\begin{proof}
Let $p\in \overline{D^+(B)}$, and let $\gamma$ be a past-inextendible causal curve through $p$. Suppose for contradiction that $\gamma \cap B =\varnothing$. Then $\gamma'=\gamma \cap C\neq \varnothing$ and $\gamma \cap \partial C =\varnothing$.  Therefore $\gamma'$ contains its future endpoint $r$. Using the submanifold structure of $C$ one can find a point $r' \in \gamma$ in a neighborhood of $r$ such that $I^-(r') \cap C \neq\varnothing$. By Wald Lemma 8.1.4~\cite{wald2010general}, there exists a timelike curve $\gamma''$ connecting $p$ to $C$. By achronality of $\Sigma$, $\gamma''$ fails to intersect $B$, which contradicts Wald Proposition 8.3.2. Hence $\gamma\cap B\neq \varnothing$, so $p\in D^+(B)$. The time-reverse of this argument shows that $D^-(B)$ is closed, and hence $D(B) = D^+(B) \cup D^-(B)$ is closed.
\end{proof}

\begin{lem}\label{lem-2}
With $B$ and $C$ as in Convention~\ref{bcconv}, $D^+(B)= D^+(\Sigma)-I^+(C)-C$ and $D(B)= M-I(C)$.
\end{lem}

\begin{proof}
$p\in D^+(\Sigma)-I^+(C)-C$ $\iff$ every past-inextendible timelike curve through $p$ intersects $\Sigma$, but none intersects $C$ $\iff$ every past-inextendible timelike curve through $p$ intersects $B$ $\iff$ $p\in D^+(B)$. The final equivalence follows from Wald Proposition 8.3.2 and Lemma \ref{lem-D+}. Combining this result with its time-reverse yields $D(B)= M-I(C)$.
\end{proof}

\begin{lem}\label{lem-3}
With $C$ as in Convention~\ref{bcconv}, $\dot{I}^+(C)-C = \dot{I}^+(\partial C) - I^{+}(C)-C$.
\end{lem}

\begin{proof}
$\dot{I}^+(\partial C)-I^+(C) -C\subseteq \dot{I}^+(C)-C$: let $q \in \dot{I}^+(\partial C)-I^+(C)-C$. If $q\in \partial C$, we are done, so we may assume from here that $q\notin \partial C$. Then since $q\in \dot{I}^+(\partial C)$, $q$ lies on a null geodesic $\gamma\subset\dot{I}^+(\partial C)$ which is either past-inextendible or has a past endpoint $p\in \partial C$~\cite{wald2010general}. Only the latter option is consistent with $\Sigma\supset \partial C$ being a Cauchy surface. Now suppose that $q\notin \dot I^+(C)$ for contradiction. Then since $q\notin I^+(C)$, there is an open neighborhood $O(q)$ that satisfies $O(q)\cap I^+(C)=\varnothing$. Hence $\gamma$ can be deformed into a timelike curve from $p$ to $r\in O(q)$; and since every open neighborhood of $p$ enters $C$, it can be further deformed to a timelike curve from $s\in C$ to $r$. This contradicts $r\notin I^+(C)$.

$\dot{I}^+(C)-C \subseteq \dot{I}^+(\partial C)-I^+(C)-C$: let $q \in \dot{I}^+(C)-C$. Again we may assume that $q\notin \partial C$ since otherwise we are done. Since $I^+(C)$ is open, $q \notin I^+(C)$. It remains to be shown that $q\in\dot I^+(\partial C)$. Note that $I^+(\partial C)\subset I^+(C)$ since every timelike curve from $\partial C$ to a point $r$ in $I^+(\partial C)$ can be deformed to a timelike curve from $C$ to $r$. Hence $q \notin I^+(\partial C)$. Now suppose that $q\notin \dot I^+(\partial C)$ for contradiction. Then there exists an open neighborhood $O(q)$ such that $I^-[O(q)]\cap \partial C=\varnothing$. Based on a similar argument to the one in the first paragraph, $q \in \dot{I}^+(C)$ implies that there exists a null geodesic $\gamma$ connecting $q$ to $C$. Let $p$ be the point of intersection with the smallest past-directed affine parameter, i.e. $\gamma$ leaves $C$ immediately to the future of $p$. Then, either $p \in \partial C$ contradicting $I^-[O(q)]\cap \partial C=\varnothing$ as in the previous paragraph. Or $p \in C$ in which case there exists a point $s$ in $\gamma$ in a neighborhood of $p$ such that $I^-(s)$ intersects $C$. This contradicts $q \in \dot{I}^+(C)$.
\end{proof}

\begin{cor}\label{lightsheet}
With $C$ as in Convention~\ref{bcconv}, let $\partial C$ be smooth. Then $\dot{I}^+(C)-C$ is the union of the future-outward directed null geodesics orthogonal to $\partial C$, terminated at either caustics or self-intersections.
\end{cor}

\begin{proof}
By the main theorem in Ref.~\cite{Akers:2017nrr}, $\dot{I}^+(\partial C)$ is the union of the future directed null geodesics orthogonal to $\partial C$, terminated at either caustics or self-intersections. By definition, the outward-directed null geodesics are those that do not enter $I^+(C)$ immediately. The result then follows from Lemma~\ref{lem-3}.
\end{proof}

\begin{prop}\label{prop-4}
With $B$ as in Convention~\ref{bcconv}, let $B'$ be a compact achronal codimension 1 submanifold such that $\partial B=\partial B'$ and $B'\subset D^+(B)$.\footnote{The latter condition can be relaxed to $B'\subset D(B)$, but this lengthens the proof and is not needed below.} Then $B$ is compact and $D(B) = D(B')$.
\end{prop}

\begin{proof}
By Wald Lemma 8.1.1, there exists a smooth timelike vector field $t^a$. Since $B' \subset D^+(B)$, the integral curves of $t^a$ that intersect $B'$ must also intersect $B$; and no curve intersects either set more than once. This provides a homeomorphism $\psi: B' \to B$ (with the image given the topology induced by $B$). Since $\partial B=\partial B'$ and $\psi(B')\subset B$, $\partial \psi(B')= \partial B$ and so $\psi(B')=B$. This implies that $B$ is compact.

Let $\gamma$ be an inextendible timelike curve, and let $p = \gamma \cap \Sigma$, so either $p\in B$ or $p\in C$. If $p \in B= \psi(B')\subset J^-(B')$, then 
$J^+(p) \cap J^-(B') \neq \varnothing$ and furthermore (by an obvious generalization of Wald Theorem 8.3.10) is compact , so $\gamma$ intersects $\dot J^-(B')$ by Wald Lemma 8.2.1. $J^-(B'-\partial B)$ is closed and so contains the closure of any of its subsets. Moreover, the closure of $B'-\partial B$ in $M$ is also its closure in $\Sigma$. Hence $B'\subset J^-(B'-\partial B)$, and $\dot J^-(B') = \dot J^-(B'-\partial B) = \dot I^-(B'-\partial B)$. Suppose that $\gamma$ did not intersect $B'-\partial B$. Then taking $C\to B'-\partial B$ in Lemma \ref{lem-3}, it follows that $\gamma$ intersects $\dot I^-(\partial B)$. This violates the Cauchy property of $\Sigma$ unless $\gamma$ intersects $\partial B$. Hence $\gamma$ must intersect $B'$, and by achronality of $B'$ it does so exactly once. If $p \in C$, it follows from $B' \subset D^+(B)$ that $\gamma \cap B' = \varnothing$. It follows that $B'\cup C$ is a Cauchy slice. By Corollary \ref{lem-2}, $D(B) = D(B')$.
\end{proof}

\begin{defn}
Let $\mu$ be a smooth compact codimension 2 submanifold of $M$. A \emph{lightsheet} $L(\mu)$ is a null hypersurface such that $\partial L\supset \mu$, with everywhere non-positive expansion away from $\mu$.
\end{defn}

\begin{conj}[Classical Bousso Bound]
The renormalized entropy on a lightsheet satisfies
\begin{equation}
    S[L(\mu)] \leq \frac{\A(\mu)}{4G }~.
\end{equation}
\end{conj}

\begin{defn}\label{ncc}
A spacetime $M$ satisfies the \emph{null curvature condition} if $R_{ab}k^a k^b\geq 0$ everywhere, where $R_{ab}$ is the Ricci tensor and $k^a$ is any null vector. (This will be the case if $M$ satisfies the Einstein equation with the null energy condition holding for the stress tensor, $T_{ab}k^ak^b\geq 0$.)
\end{defn}

\begin{thm}[Singularity Theorem for Hyperentropic Regions]
\label{cst}
Let $B$ be a closed subset of the Cauchy slice $\Sigma$ of $M$, with smooth compact boundary $\partial B$ as in Convention~\ref{bcconv}. Let the null curvature condition and the Bousso bound hold on $M$. If the future-directed inward null geodesic congruence orthogonal to $\partial B$ has negative expansion, and moreover if $S(B)>\A(\partial B)/4G$, then at least one of the above null geodesics is incomplete.
\end{thm}

\begin{proof}
Let $L=\dot I^+(C)-C$. By Wald Theorem 8.1.3, $\dot I^+(C)$ is a codimension 1 embedded $C^0$ achronal submanifold of $M$. Hence $L$ is a submanifold with boundary $\partial L=\partial B$, and by corollary~\ref{lightsheet}, $L$ is a null hypersurface. By assumption, the expansion $\theta$ of $L$ is negative on $\partial B$, and by the null curvature condition it will remain negative everywhere on $L$, so $L$ is a lightsheet.

By compactness of $\partial B$, $\theta$ is bounded away from 0; hence by rescaling the affine parameter at different points, we can arrange that $\theta=-1$ everywhere on $\partial B$. By the focussing theorem of general relativity, all generators of $L$ must leave $L$ at or before affine distance $2$ from $\partial B$. Assuming for contradiction that every generator is complete, then $L$ is a subset of the union ${\cal L}$ of the null geodesics orthogonal to $\partial B$ in the closed interval $[0,2]$. Since $\partial B$ is compact, then so is ${\cal L}$; and since $L$ is closed, $L$ must be compact.

By Lemma \ref{lem-3}, no point in $L$ can be connected by a past-inextendible timelike curve to $C$. Therefore, by the Cauchy property of $\Sigma$, all past-inextendible timelike curves passing through $L$ must intersect $B$. Hence $L \subset \overline{D^+(B)}$, and by Lemma \ref{lem-D+}, $L \subset D^+(B)$. Since $\partial L=\partial B$, Proposition \ref{prop-4} implies that $B$ is compact and that $D(L)=D(B)$. Thus, by unitarity, $S(L)=S(B)$. This contradicts the Bousso bound, so at least one null generator of $L$ must be incomplete.
\end{proof}

\begin{figure}
\includegraphics[width=0.3\textwidth]{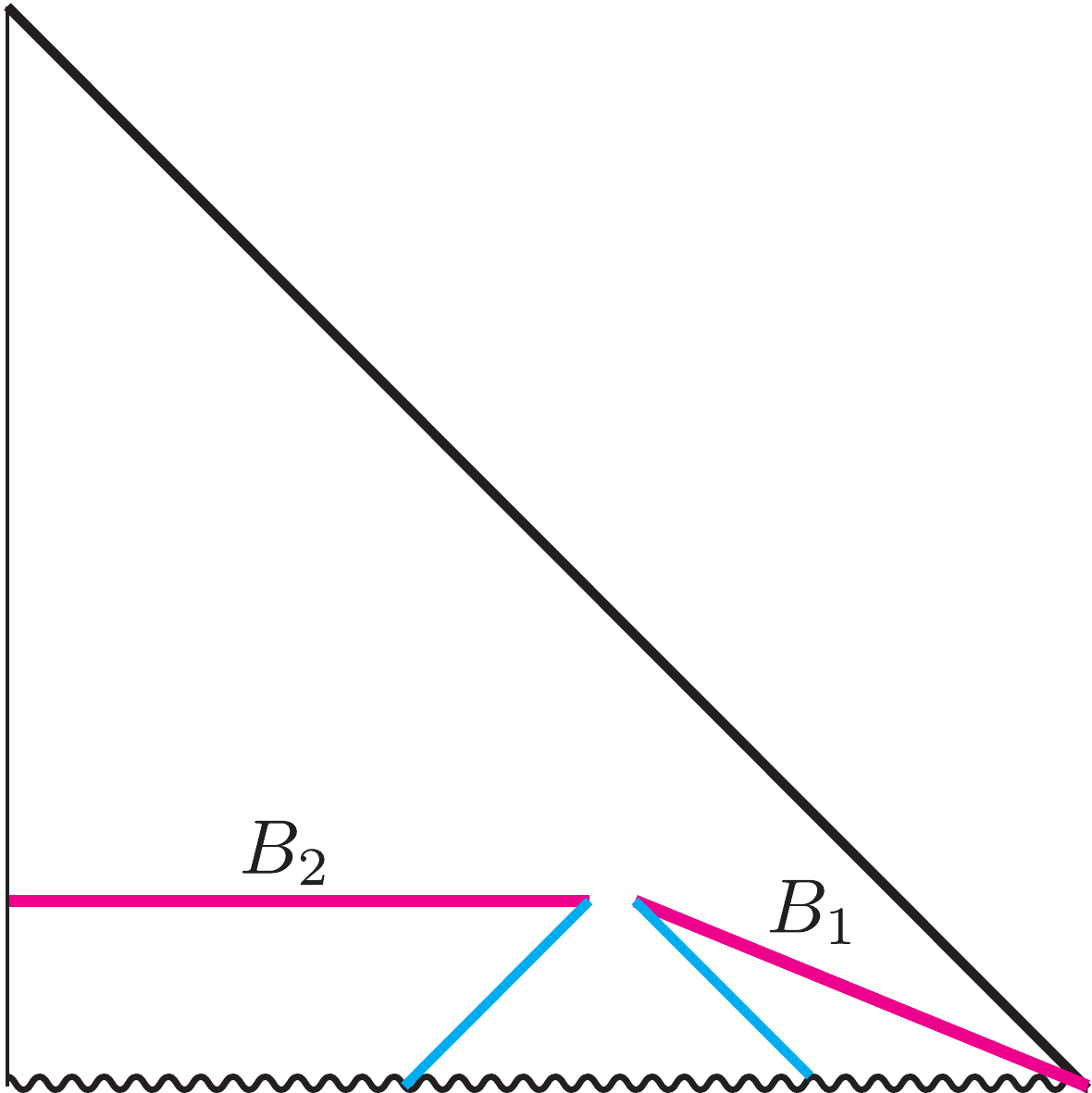}
\caption{\label{Figure_FRW}
{\footnotesize Spatially flat expanding universe with pressureless dust. Both the noncompact region $B_1$ and the compact region $B_2$ are hyperentropic, and the past-directed lightsheets entering them run into the big bang singularity.}}
\end{figure}

\section{Discussion} \label{sec-dis}

Our theorem, like Penrose's, provides sufficient conditions for a singularity, but not an equivalence. For example, neither theorem predicts a singularity from the data on an early Cauchy surface of a collapsing dust cloud that eventually forms a black hole. 

Our theorem does not require the fine-grained entropy to be large. One can coarse-grain the matter entropy in $B$ while holding fixed the stress tensor and thus the spacetime geometry. If the coarse-grained state is hyperentropic, then there will be a singularity even if the fine-grained entropy in $B$ is small.

Both Penrose’s theorem and ours are proven by demonstrating that $B$ cannot evolve to a compact null hypersurface $L$. In Penrose’s case, the obstruction is topological, since $B$ is noncompact. In our theorem, the obstruction is unitarity: if all the entropy on $B$ passed through $L$, then the Bousso bound would be violated.

Yet, in a broad class of examples, our theorem implies Penrose's. Suppose for example that $B$ is asymptotically flat and satisfies the conditions of Penrose’s theorem. If $B$ is not hyperentropic, then it would appear that our theorem does not apply. But in the asymptotic region of $B$, one can add soft radiation carrying an arbitrary amount of entropy with arbitrarily small backreaction. Now our theorem does imply a singularity, and since the metric is unchanged, the singularity must have been present in the original spacetime. This argument is not fully general, because it is not clear that every noncompact $B$ can be made hyperentropic with small backreaction in $D(B)$. But it gives us a heuristic understanding of the relation between the theorems.

Penrose's theorem gives no information about the character of the singularity it predicts, and our theorem shares this limitation. Geodesic incompleteness need not be accompanied by the divergence of a curvature invariant. For example, it could be caused by a violation of strong cosmic censorship: for a complete initial data slice $\Sigma$, $M=D(\Sigma)$ is a proper subset of a maximally extended spacetime $\mathbf{M}$. In that case, let  $H^+(\Sigma)$ denote the future Cauchy horizon of $\Sigma$ in $\mathbf{M}$. Since $\Sigma$ is complete, it has no edge, so by Wald Theorem 8.3.5, every point $p\in H^+(\Sigma)$ lies on a past-inextendible null geodesic contained entirely in $H^+(\Sigma)$. Hence $H^+(\Sigma)$ either contains an incomplete null geodesic or a past asymptotic region. If $p$ lies also in the closure of $D^+(B)$ in $\mathbf{M}$, then the past-inextendible geodesic is contained entirely in this closure.

\section{Examples} \label{sec-ex}

Consider an expanding Friedmann-Robertson-Walker universe whose energy density is dominated by pressureless dust (Fig.~\ref{Figure_FRW}). At any fixed time, sufficiently large spheres are anti-trapped~\cite{Bousso:1999xy}; let $B_1$ be the exterior of such a sphere. The entropy $S(B)$ is infinite, so both conditions of our theorem are satisfied (with past and future exchanged). The Penrose theorem, too, would have predicted the big bang singularity from data on $B_1$, since $B_1$ is noncompact.

However, Theorem~\ref{cst} can diagnose the big bang from compact regions as well. In the above example, the past-inward directed null geodesics orthogonal to any sphere have negative expansion. At any fixed time, the entropy in a ball $B_2$ of proper radius $r$ scales as $r^3$, whereas the area scales as $r^2$. Hence, for large enough radius, $B_2$ will be hyperentropic.

\begin{figure}
%    \centering
\includegraphics[width=0.3\textwidth]{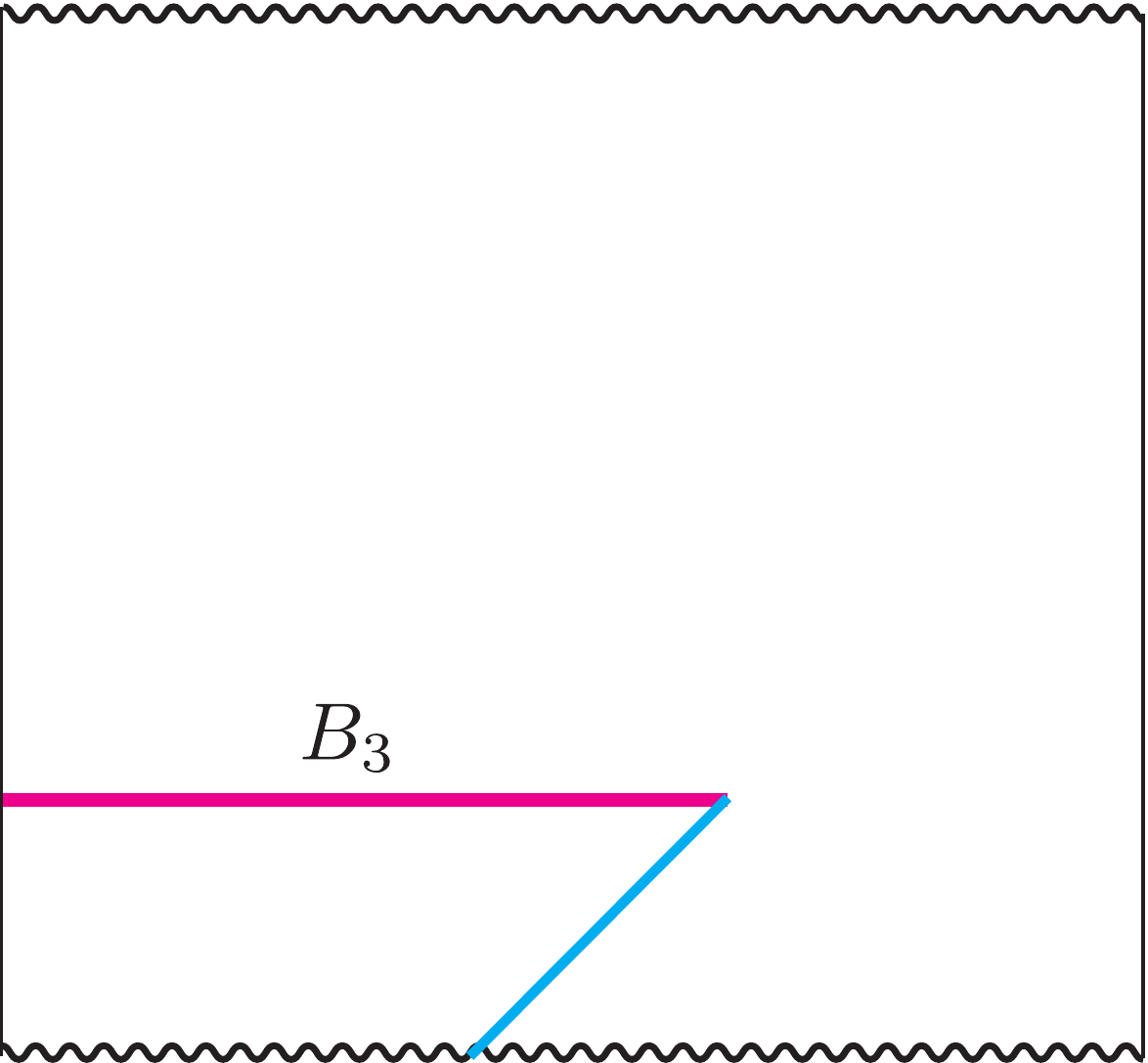}
\caption{\label{Figure_closed1}{\footnotesize
Closed radiation-dominated universe. The compact hyperentropic region $B_3$ is similar to $B_2$ in Figure~\ref{Figure_FRW}.}}
\end{figure}

The Penrose theorem never applies in a closed universe, since all spatial regions are compact, but our Theorem~\ref{cst} does. A simple example is the region $B_3$ in Fig.~\ref{Figure_closed1}, obtained by introducing positive spatial curvature on a scale much greater than the hyperentropic region $B_2$. 

\begin{figure}
%    \centering
\includegraphics[width=0.3\textwidth]{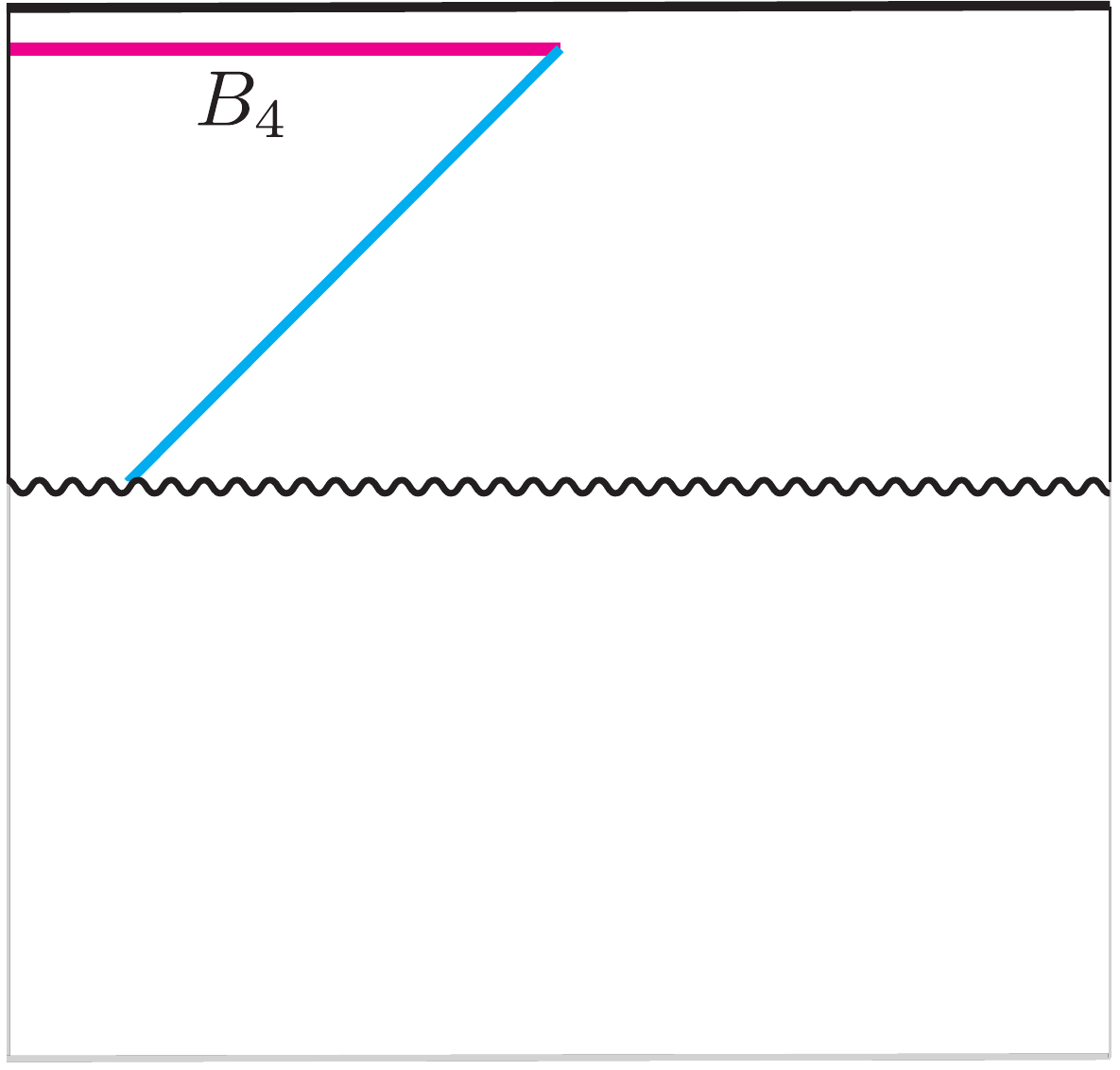}
\caption{\label{Figure_closed2}{\footnotesize
The full square without the wiggly line represents de Sitter space, which is nonsingular. A late hemisphere $B_4$ can be made hyperentropic by adding soft radiation with negligible local backreaction. Under backward evolution this results in a big bang, as predicted by our theorem.}}
\end{figure}

To showcase the power of Theorem~\ref{cst}, let us study a closed universe with cosmological constant $\Lambda>0$  (Fig.~\ref{Figure_closed2}). Our construction begins with the vacuum de Sitter solution, which is not singular. Each time slice is a three-sphere of radius $a(t)\propto t_\Lambda \cosh(t/t_\Lambda)$, where $t_\Lambda=(3/\Lambda)^{1/2}$. Let $\Sigma$ be the sphere at a a very late time $t\gg t_\Lambda$, and modify the data on $\Sigma$ by adding thermal radiation with negligible density $\rho\ll \Lambda/G$. This will not change the metric much near $\Sigma$, but it can make the left hemisphere $B_4$ hyperentropic. The radiation entropy in the left hemisphere $B_4$ is $S_{B_4}\sim \rho^{3/4} a^3$. The equator has area $A(\partial B_4)\sim a^2$, and its past-directed orthogonal light rays have negative expansion. Thus, the conditions of Theorem~\ref{cst} can be satisfied by choosing $a$ large at fixed $\rho$. Note that the universe can be made to look arbitrarily close to empty de Sitter space for as much time as we like before $\Sigma$, by choosing $a$ large and $\rho$ small. Yet, explicit calculation shows that if the late hemisphere is hyperentropic, the radiation density will eventually come to dominate over the vacuum energy under backward time evolution, replacing the bounce with a big bang. Thus, the Theorem diagnoses a big bang in the arbitrarily distant past.

\begin{figure}
%    \centering
\includegraphics[width=0.3\textwidth]{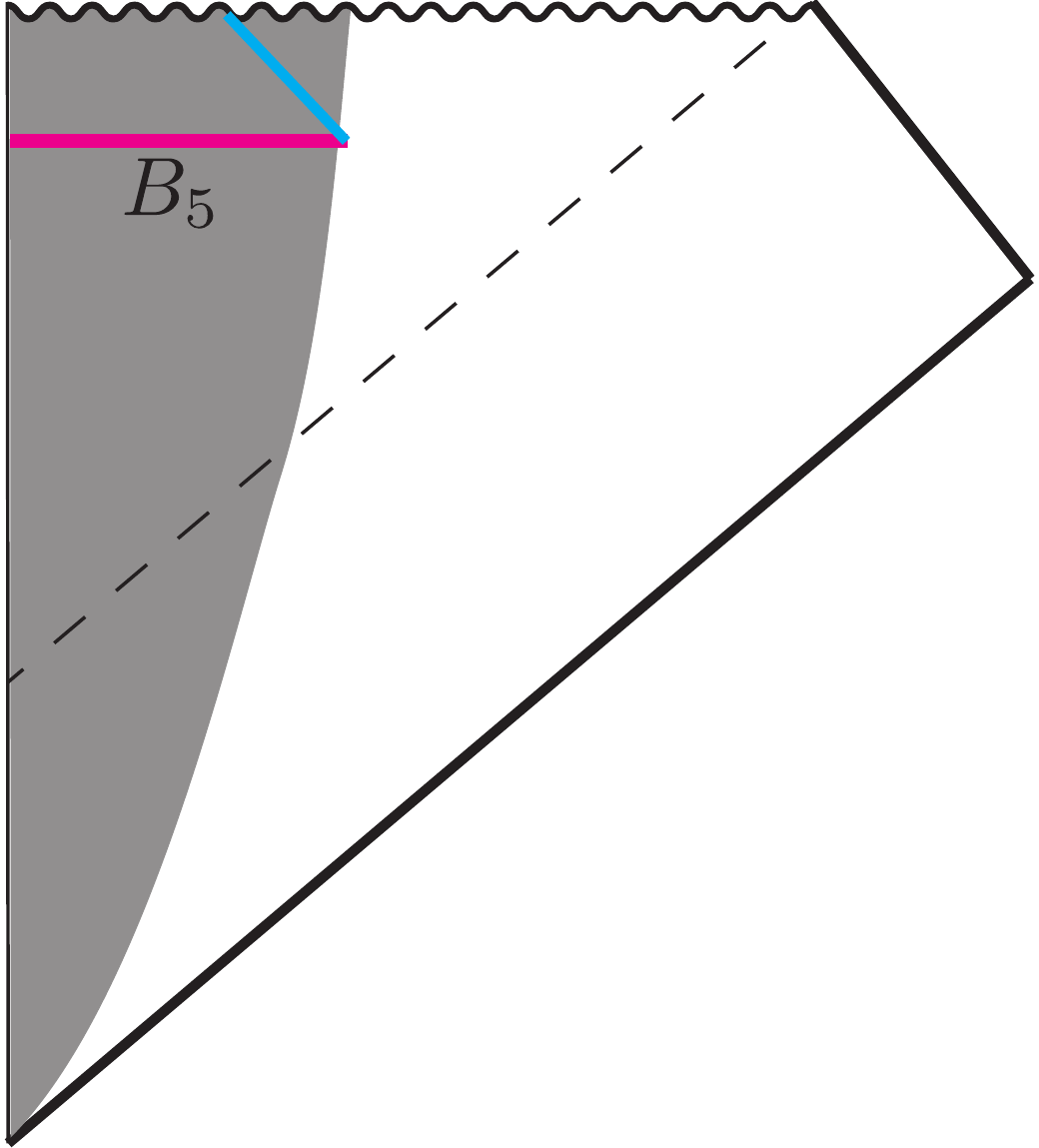}
\caption{\label{Figure OppSny}{\footnotesize
Gravitational collapse of a dust ball (shaded). The lightsheet of the hyperentropic region $B_5$ runs into the future black hole singularity.}}
\end{figure}

Finally, consider the gravitational collapse of a ball of pressureless dust (a ``star'') of mass $M$ shown in Fig.~\ref{Figure OppSny}. This can be constructed by cutting out the hyperentropic region $B_2$ in Fig.~\ref{Figure_FRW} and reversing time~\cite{Oppenheimer:1939ue}. The exterior is a vacuum solution, shown here as noncompact, but our theorem applies even if the global topology is compact.

\paragraph{Acknowledgements}
We would like to thank D.~Harlow, G.~Horowitz and R.~Mahajan for discussions. We are grateful to A.~Chatwin-Davies, C.~Murdia, A.~Rolph, and E.~Shaghoulian for comments on Ref.~\cite{Bousso:2021sji} that indirectly~\cite{QuantumTheorem} motivated this paper.  This work was supported in part by the Berkeley Center for Theoretical Physics; by the Department of Energy, Office of Science, Office of High Energy Physics under QuantISED Award DE-SC0019380 and under contract DE-AC02-05CH11231; and by the National Science Foundation under Award Numbers 2112880 (RB) and 2014215 (ASM).

\bibliographystyle{utcaps}
\bibliography{main}

\end{document}